\documentclass[sigconf]{acmart}
\AtBeginDocument{%
  }

\copyrightyear{2026}
\acmYear{2026}
\setcopyright{cc}
\setcctype{by}
\acmConference[AISec '26]{19th Workshop on Artificial Intelligence and Security}{November 15--19, 2026}{The Hague, Netherlands}
\acmBooktitle{19th Workshop on Artificial Intelligence and Security (AISec '26), November 15--19, 2026, The Hague, Netherlands}
\acmDOI{10.1145/3847352.3848113}
\acmISBN{979-8-4007-3031-3/2026/11}

\usepackage{booktabs}
\usepackage{multirow}
\usepackage{graphicx}
\usepackage{amsmath}

\usepackage{amssymb}
\usepackage{amsthm}
\usepackage{xcolor}
\usepackage{tikz}
\usepackage{pgfplots}
\pgfplotsset{compat=1.18}
\usepackage{enumitem}
\usepackage[ruled,vlined]{algorithm2e}

\newtheorem{definition}{Definition}

\begin{document}

\title{
TERMon: Detecting Persistent Behavioral Threats in Edge AI via Hardware-Native Ternary Runtime Monitor
}

\author{Arish Sateesan}
\orcid{0000-0002-8197-0097}
\affiliation{%
  \institution{Aalborg University}
  \city{Copenhagen}
  \country{Denmark}}
\email{arishs@es.aau.dk}

\author{Edlira Dushku}
\orcid{0000-0002-4974-9739}
\affiliation{%
  \institution{Aalborg University}
  \city{Copenhagen}
  \country{Denmark}}
\email{edu@es.aau.dk}

\renewcommand{\shortauthors}{Arish Sateesan and Edlira Dushku}

\begin{abstract}
Edge AI accelerators are increasingly deployed in safety-critical environments, where model outputs may control physical actuators, make access-control decisions, or trigger alarms. In these settings, runtime failures often remain undetected because model corruption, distribution shift, and adversarial inputs can still produce well-formed, confident predictions. This paper presents TERMon, a lightweight hardware runtime monitor that detects such anomalies by observing inference behavior rather than re-executing or formally verifying the model. 
TERMon represents class-conditional trusted behavior as hardware-efficient ternary patterns that are matched in parallel against a thermometer-encoded fingerprint. The ternary encoding reproduces the corresponding unquantized range decision exactly. TERMon detects harmful weight corruptions in proportion to their behavioral impact, while out-of-distribution and adversarial inputs are largely not separable using the monitored features at a strict false-positive operating point. We implemented TERMon on a PYNQ-Z2 FPGA, and the pipelined design requires no on-chip block RAM or DSPs and has a two-cycle decision latency.
\end{abstract}
 





\maketitle

\section{Introduction} \label{label:intro}

Machine-learning~(ML) models are increasingly deployed at the edge to perform inference across industrial sensing, autonomous systems, medical devices, space systems, and smart infrastructure~\cite{zhou2019edgeintelligence,Furano2020aispace,Kuutti2021dlautonomousvehicle}. Unlike models used for retrospective data analysis, these models produce outputs that can directly control actuators, determine access, or trigger alarms. An incorrect prediction issued with high confidence can therefore lead to a safety or security incident. The main challenge is that many runtime failures produce no explicit execution error: the model continues to return syntactically valid, often high-confidence outputs even when the computation, the input distribution, or the input itself should no longer be trusted~\cite{Guo2017oncalibration,Ovadia2019canyoutrust}.

\begin{sloppypar}
This silent-failure behaviour spans several threat classes. Physically exposed edge hardware is vulnerable to memory faults and fault-injection attacks that corrupt stored model weights while leaving the surrounding software stack apparently intact~\cite{kim2014rowhammer,rakin2019bitflip}. Deployed sensors may age, degrade, or drift outside their intended operating conditions, causing the model to receive out-of-distribution~(OOD) inputs to which deep neural networks~(DNN) may assign unjustified confidence~\cite{hendrycks2017baseline, nguyen2015deep}. Adversarial examples pose a further concern, as carefully crafted inputs can change the model decision while remaining close to benign inputs~\cite{goodfellow2014explaining, madry2017towards}.
\end{sloppypar}

Existing defenses address parts of this problem, but they differ in their assumptions about what must be checked and when a system should respond. Weight-integrity checks are well suited when every modification of stored weights must be detected. However, in deployments where recovery is costly, such as space systems or remote sensing platforms, halting execution or restoring the model after every detected bit flip may be unnecessarily expensive when the corruption does not affect inference behavior. Software anomaly and OOD detectors can examine detailed runtime model behavior, but they are less isolated from the software stack they monitor and cannot guarantee a fixed hardware-level response. Their response may therefore be delayed by system load or disrupted if the software is compromised. A hardware monitor can instead provide deterministic, fixed-latency response independently of software scheduling. This motivates a lightweight hardware mechanism that checks whether the model's observed inference behavior remains within the ranges learned from an uncorrupted model. 

\begin{sloppypar}
This paper proposes \emph{TERMon}, a Ternary Envelope Runtime \mbox{Monitor} that detects runtime anomalies by observing inference behavior rather than re-executing or formally verifying the model. TERMon monitors eight scalar features extracted from the input, intermediate activations, and output distribution, and checks them against ranges learned from clean inferences. A clean inference denotes an inference performed with an uncorrupted model on an in-distribution, non-adversarial input. TERMon does not identify the threat type or correct the model. Instead, it flags inferences whose monitored behavior deviates from the learned clean ranges.
\end{sloppypar}

TERMon is designed for lightweight hardware implementation. Each feature is quantized, and thermometer encoded so that its trusted range maps to a ternary pattern with fixed and \emph{don't-care} positions. This allows the runtime decision to be implemented as a compact ternary content-addressable memory~(TCAM)-compatible match. For persistent threats, such as corrupted weights or sensor drift, TERMon aggregates per-inference flags using a short sequential rule~(cf. Section~\ref{subsec:sequential_deci}). 

Our evaluation on the CIFAR-10 dataset~\cite{krizhevsky2009learning} leads to three main findings.
First, TERMon exhibits \emph{impact-proportional} detection: benign corruptions are flagged near the false-positive rate measured on clean inferences~(clean false-positive rate), whereas accuracy-degrading corruptions are detected reliably over consecutive inferences. Second, the hardware-compatible ternary representation does not limit weight-fault detection. Third, the monitored features are insufficient for reliable per-input detection of OOD inputs or successful FGSM and PGD adversarial examples at a low false-positive rate. 

This paper makes the following contributions:
\begin{itemize}[topsep=2pt, partopsep=0pt, leftmargin=4mm]
    \item We propose TERMon, a lightweight hardware runtime monitor that checks inference-level behavioral features against learned clean ranges using a TCAM-compatible ternary representation.
    \item We introduce an evaluation methodology that separates hardware-encoding limits from feature-representation limits, and use it to derive practical design choices for behavioral monitors.
    \item We implement TERMon on a PYNQ-Z2 FPGA and report post-implementation resource utilization, timing, and power.
\end{itemize}

\section{Background} \label{sec:background}
\subsection{Runtime Threats to Edge AI} \label{subsec:runtime_threats}
Edge AI systems can fail at runtime while still producing well-formed and confident predictions. This makes runtime monitoring important, especially when the model output is used for actuation, access control, or safety-related decisions.

\textit{Fault injection.}
Fault injection can corrupt either the model state or the hardware that executes the model. For example, dynamic random-access memory~(DRAM) disturbance errors can be induced deliberately without direct software access to the target's data~\cite{kim2014rowhammer}, and targeted bit-flip attacks show that modifying a small number of weight bits can severely degrade the accuracy of the DNN model~\cite{rakin2019bitflip}. Even non-adversarial bit flips can propagate through DNN accelerators in ways that depend strongly on the affected layer, data type, and bit position~\cite{li2017duplication}. In floating point weights, exponent-bit flips can be especially damaging because they may turn a normal weight into an abnormally large value. Existing countermeasures focus mainly on redundancy, replication, or integrity checks, and therefore protect the stored data or computation rather than checking whether the resulting inference behavior remains normal.

\textit{Out-of-distribution inputs.}
OOD inputs occur when a deployed model receives inputs that differ from the distribution observed during training. This can happen because the environment changes, a sensor degrades, or the deployed system observes data that the model was not trained to handle. DNNs can assign high confidence to such inputs, including inputs far outside the training distribution~\cite{hendrycks2017baseline, nguyen2015deep}. Software OOD detectors based on confidence scores, calibration, or layer-wise Gaussian statistics can improve detection on suitable models~\cite{lee2018mahalanobis}. However, many reported results assume false-positive rates well above the 1\%, which is higher than a safety-critical edge deployment can tolerate. In this work, we evaluate OOD behavior at a strict 1\% false-positive operating point.

\textit{Adversarial examples.}
Adversarial examples are inputs deliberately crafted to induce misclassification while appearing benign~\cite{goodfellow2014explaining, madry2017towards}. A substantial body of work has focused on detecting such inputs, while adaptive attacks have repeatedly shown that many proposed detectors can be bypassed once the attacker knows the decision rule~\cite{xu2017feature, carlini2017detected}. Hence, this work does not claim adversarial robustness. Instead, we include adversarial examples to assess whether the monitored behavioral features contain enough information to distinguish successful attacks from clean inferences.
%
\subsection{Hardware Runtime Monitoring} \label{subsec:hw_monitoring}
Hardware runtime monitors are advantageous for edge AI because they offer deterministic latency, bounded overhead, and isolation from the software stack running the model. Existing hardware defenses for ML accelerators are typically designed around specific threat models, such as redundancy-based fault tolerance~\cite{zhang2018faultaware} and integrity checks that verify model state or data movement~\cite{li2017duplication}. In contrast, behavioral anomaly detection is usually implemented in software, where it is easier to deploy but less isolated from the system being monitored.
This work studies hardware behavioral monitoring as a lightweight alternative. Rather than checking one specific fault mechanism, the monitor observes whether the model's runtime behavior remains within ranges learned from clean inferences, which separates the mechanism from the threat source. We position the monitor against existing runtime defenses in  Section~\ref{sec:related_work}.

\subsection{Ternary Matching} \label{subsec:cam}
A content-addressable memory~(CAM) compares a binary input query against all stored entries in parallel and returns a match in constant time~\cite{pagiamtzis2006cam}. A ternary CAM~(TCAM) extends the functionality by allowing each stored bit to take one of three values: 0, 1, or \emph{don't care}. Ternary matching can therefore compactly represent partial or masked patterns and ranges. TERMon uses this ternary matching functionality to represent trusted feature ranges.
Unlike conventional TCAM applications that store large rule tables or associative databases, TERMon needs only ten ternary entries~(cf. Section~\ref{subsec:trust_envelope}) over a small number of monitored features. A ternary pattern can be represented as a value register together with a care-mask register that marks which positions must match, so the ternary match reduces to a masked comparison in ordinary logic. Our implementation therefore uses TCAM-compatible value and care-mask registers rather than a large TCAM memory array. Hence, improving memory efficiency and scalability of TCAMs for large keys and large entry sets, which are known hardware limitations that recent works address~\cite{sateesan2026breaking}, are complementary to our setting.

\section{Threat Model and Security Objective} \label{sec:threatmodel}
In this section, we define the runtime threat model for Edge AI, describe the trust assumptions, and specify the security objectives of a hardware runtime monitor for detecting behavioral anomalies in edge-AI inference.

\subsection{Threat model}
We consider runtime attacks against Edge AI systems that alter the deployed model state or manipulate inference inputs while the model continues to operate and produce well-formed predictions. Our adversarial scope covers fault injection and adversarial examples. We also consider out-of-distribution inputs as a non-adversarial runtime condition that may lead to unreliable predictions.

\par \textbf{Fault injection.} We consider an attacker capable of inducing bit flips in memory locations storing the model's weights~\cite{kim2014rowhammer, rakin2019bitflip}. We evaluate this threat for both 32-bit floating-point~(FP32) and 8-bit floating-point~(FP8) weight representations. Such faults may be induced through physical fault-injection techniques or software-induced DRAM disturbance attacks, as discussed in Section \ref{subsec:runtime_threats}. We abstract away the fault-delivery mechanism and consider its effect on the stored weight values. When an induced bit flip changes a stored weight, the corrupted weight remains in use across subsequent inferences until it is restored or the model is reloaded. Similar weight corruptions may also result from non-adversarial hardware faults~\cite{li2017duplication}.

\par \textbf{Out-of-distribution inputs.} We consider inference inputs drawn from a distribution different from that represented during model training. Such inputs may arise from changes in the operating environment, sensor degradation, or data outside the model’s intended domain~\cite{hendrycks2017baseline, Ovadia2019canyoutrust}. We treat OOD inputs as a non-adversarial runtime condition rather than an attacker action.

\textbf{Adversarial examples.} We consider an attacker with white-box knowledge of the model, including its architecture and parameters, who can modify inference inputs to cause misclassification~\cite{goodfellow2014explaining, madry2017towards}. The attacker uses this knowledge when crafting the inputs but does not use knowledge of TERMon’s monitored features or decision rule.

\par \textit{Assumptions.} Our threat model excludes faults affecting activations, registers, arithmetic units, control flow, or feature extraction. We also exclude attacks that tamper with the features delivered to TERMon, alter or disable the monitor, or prevent the model from producing a prediction. The adversarial-example attacker does not use knowledge of TERMon when crafting inputs; attacks designed to cause misclassification while evading TERMon are outside scope.

\begin{figure}[t]
\centering
\includegraphics[width=0.8\columnwidth]{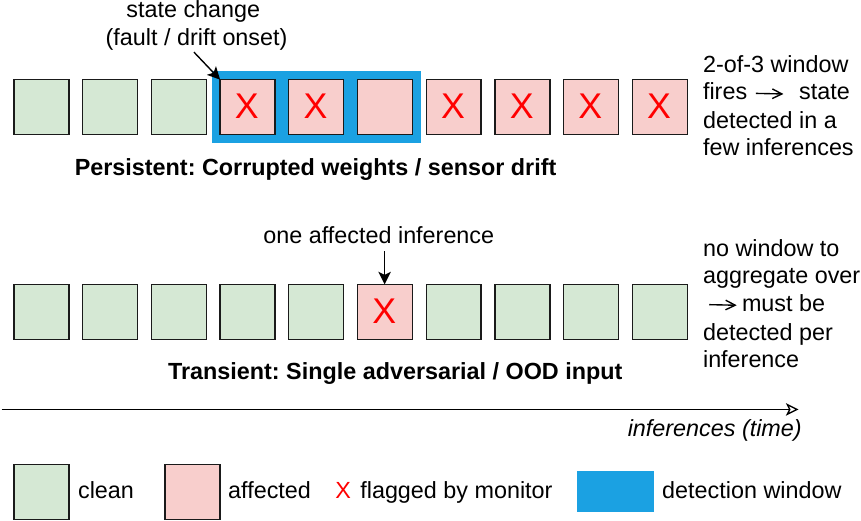}
\caption{Persistent threats (top) affect every subsequent inference and are
detectable by sequential aggregation over a window; transient threats (bottom)
affect a single inference and must be caught on that inference alone.
\vspace{-2mm}
}
\label{fig:threat}
\end{figure}

\subsubsection{Persistent and Transient anomalies}
We distinguish runtime anomalies according to how long the underlying cause remains active, as shown in Figure~\ref{fig:threat}.
A \textbf{persistent anomaly} changes the deployed model state or the input-generation process and remains active across multiple inferences. Examples include corrupted model weights and continued sensor degradation that causes a persistent change in the input distribution. A persistent anomaly may remain active across multiple inferences without causing every inference to be misclassified or flagged.
A \textbf{transient anomaly} affects a single inference and leaves no lasting change in the deployed model state or input-generation process. Examples include an isolated OOD input and a single adversarial example.
This distinction determines the evidence available to TERMon. A transient threat provides one opportunity for detection and must therefore be identified from the affected inference. A persistent anomaly provides repeated opportunities for detection, allowing TERMon to aggregate per-inference flags using the sequential rule described in Section~\ref{subsec:sequential_deci}.

\subsection {Security objective}
Based on the distinction between persistent and transient runtime anomalies, the designed monitor, TERMon, has two detection objectives.
First, at the inference level, it should flag inferences whose behavioral features leave the learned clean ranges while keeping the false-positive rate on clean inferences within a specified budget. Second, for persistent threats, it should detect the underlying abnormal state within a small number of inferences while maintaining a bounded false-alarm probability.

\section{TERMon Design} \label{sec:monitor_design}
TERMon performs runtime detection in two stages: a per-inference behavioral check and temporal aggregation for persistent anomalies. For each inference, it monitors eight scalar features from the input, intermediate activations, and output distribution. Trusted ranges for these features are learned offline and encoded as class-specific ternary patterns. At runtime, the feature values are quantized and thermometer encoded, then matched against the ternary patterns. The result corresponding to the predicted class determines the per-inference flag. The sequential rule aggregates these flags to detect persistent anomalous states.
\subsection{Monitored Features} \label{sec:monit_features}
We consider an AI model executing on an embedded computing platform, where TERMon passively monitors its inference behavior. The goal is to capture changes in runtime behavior using a small number of scalar features taken from different stages of the inference pipeline. We therefore monitor the input, intermediate activations, and output distribution, capturing the changes in the incoming data, internal computation, and the model's confidence and decision behavior. We use eight features that provide this coverage while remaining simple to compute in hardware. Section~\ref{subsec:lessons} discusses this design rationale.

Table~\ref{tab:features} summarizes the eight monitored features and their tap points. The input features $f_1$ and $f_2$ capture changes in the data presented to the model. The activation features $f_3$-$f_5$ capture changes in internal computation, including unusually large activation values caused by corrupted weights. The output features $f_6$ to $f_8$ summarize the model's confidence profile.
Each monitored feature is a scalar summary of the tensor or output vector computed at its tap point without storing the full tensor or re-running the model. For tensor-based features, the values are processed in a single streaming pass. Let $z_t$ be the scalar value arriving at clock cycle $t$. A state register is updated as $s \leftarrow g(s,z_t)$, where the update function $g$ depends on the feature. For $f_1$-$f_3$, it computes running sums, $s \leftarrow s + z_t$, with an additional sum of squares for the standard deviation $f_2$. For $f_4$, it counts zeros, $s \leftarrow s + \mathbf{1}[z_t = 0]$, to measure sparsity. For $f_5$, it tracks the maximum activation, $s \leftarrow \max(s, z_t)$. The output features are computed directly from the logit vector. The feature $f_6$ is the top-1 softmax probability, $f_7$ is the top-1--top-2 margin, and $f_8$ is the softmax entropy. Thus, the monitored features require only streaming operations such as accumulation, zero counting, maximum tracking, and output comparison, making them suitable for small fixed-point circuits at their tap points.

\begin{table}[t]
\centering
\caption{Behavioral features and tap points.}
\vspace{-3mm}
\label{tab:features}
\resizebox{0.65\linewidth}{!}{%
\begin{tabular}{cll}
\toprule
\textbf{Feature} & \textbf{data} & \textbf{Tap point} \\
\midrule
$f_1$ & Mean of input tensor         & Input  \\
$f_2$ & Std. deviation of input tensor        & Input  \\
$f_3$ & Mean post-ReLU activation    & Conv1  \\
$f_4$ & Sparsity (fraction of zeros) & Conv1  \\
$f_5$ & Max post-ReLU activation     & FC1    \\
$f_6$ & Top-1 softmax probability    & Output \\
$f_7$ & Margin (top-1 $-$ top-2)     & Output \\
$f_8$ & Output entropy               & Output \\
\bottomrule
\end{tabular}
}
\end{table}

\subsection{Trusted Behavioral Envelope} \label{subsec:trust_envelope}
\begin{definition}[Trusted Behavioral Envelope] \label{def:envelope}
For predicted class $c$ and feature $j$, let $[\ell_{c,j}, h_{c,j}]$ denote the $\bigl(\tfrac{\alpha}{2d},\, 1-\tfrac{\alpha}{2d}\bigr)$ quantiles of feature $j$ over the clean inferences that the model assigns to class $c$, where $\alpha$ is the false-positive budget and $d=8$ is the number of monitored features. An inference with feature vector $(x_1,\dots,x_d)$ and predicted class $c$ is \emph{trusted} iff $x_j \in [\ell_{c,j}, h_{c,j}]$ for every feature $j$. Otherwise the inference is flagged.
\end{definition}
We refer to these per-feature trusted ranges collectively as a behavioral envelope because they define the bounds within which clean inference behavior is expected to remain.
The envelope is conditioned on the model's own prediction, which constrains an inference to behavior typical of that class. We also evaluate a \emph{global} variant, in which the same quantiles are taken over all clean inferences irrespective of class, giving a single envelope and one stored entry rather than~$C$. The global variant cannot detect a mismatch between the predicted class and the observed behavior. An inference whose features appear normal across all clean inferences, but atypical for its predicted class may leave the class-conditional envelope while remaining inside the global envelope. We retain the global single-envelope as a baseline in Section~\ref{subsec:rq1_fi}.
The ranges $[\ell_{c,j}, h_{c,j}]$ describe the behavior of clean inferences rather than identifying specific training inputs. A previously unseen but in-distribution input is therefore trusted whenever its features remain within the ranges learned from clean inferences. Thus, the envelope is a behavioral region, not an input whitelist. Non-adversarial OOD inputs are considered separately in Section~\ref{subsec:limits}.

The envelope applies an independent range check to each feature rather than combining all deviations into a summed anomaly score. This corresponds to an $L_\infty$-style, or maximum-deviation, decision rule, i.e., an inference is flagged when any monitored feature leaves its trusted range. This choice is important because the anomalies studied here are often concentrated in a small number of features. For example, corrupted weights mainly produce unusually large internal activation values, whereas severe contrast corruption mainly affects the input standard deviation. A summed score can dilute such deviations by averaging one abnormal feature with several features that remain close to normal. Section~\ref{subsec:lessons} compares this design with summed-distance alternatives. The false-positive budget $\alpha$ determines the width of the trusted ranges. Each feature receives a probability budget of $\alpha/d$, split equally between the lower and upper tails. A smaller $\alpha$ therefore produces wider ranges and fewer false positives, whereas a larger $\alpha$ produces narrower ranges and flags more inferences.
%
\subsection{Range-Aligned Thermometer Encoding} \label{subsec:encoding}
Definition~\ref{def:envelope} yields one envelope per class, defined by one trusted interval for each of the $d$ monitored features. Rather than evaluating them numerically at runtime, we encode the features once against a threshold bank shared by all classes and represent each envelope as a single ternary pattern. The resulting fingerprint~($\texttt{fp}$) is matched against all class-specific patterns in parallel.
\begin{equation}
{T}_j = \operatorname{sort} \bigl(\ell_{1,j}-\delta,h_{1,j},\ldots, \ell_{C,j}-\delta,h_{C,j}\bigr),
\label{eq:bank}
\end{equation}
Here, $\delta$ denotes one quantization step of the feature representation. TERMon does not require a particular fixed-point scaling, provided the feature values and stored thresholds use the same signed 16-bit representation. To preserve the inclusive lower bound in Definition~\ref{def:envelope}, the offline configuration generator computes the lower threshold as $\ell_{c,j}-\delta$. Thus, the hardware comparison $x_j>\ell_{c,j}-\delta$ is equivalent to $x_j\ge\ell_{c,j}$ on the fixed-point grid. The hardware stores only these resulting thresholds. The bank contains at most $2C$ distinct values, and when bounds coincide across classes, threshold values may be repeated in the fixed $2C$ hardware positions. With $C=10$, each feature therefore has 20 threshold positions and the eight codes form an $8\times20=160$-bit fingerprint.
At runtime, the comparators for feature $j$ evaluate $\texttt{fp}_{j,k} = \mathbf{1}[x_j > {T}_{j,k}]$. Since ${T}_j$ is sorted, the outputs form a thermometer code and no separate encoder is required. Concatenating the $d$ codes gives the fingerprint $\texttt{fp}$.

\begin{figure}[t]
\centering
\includegraphics[width=1\columnwidth]{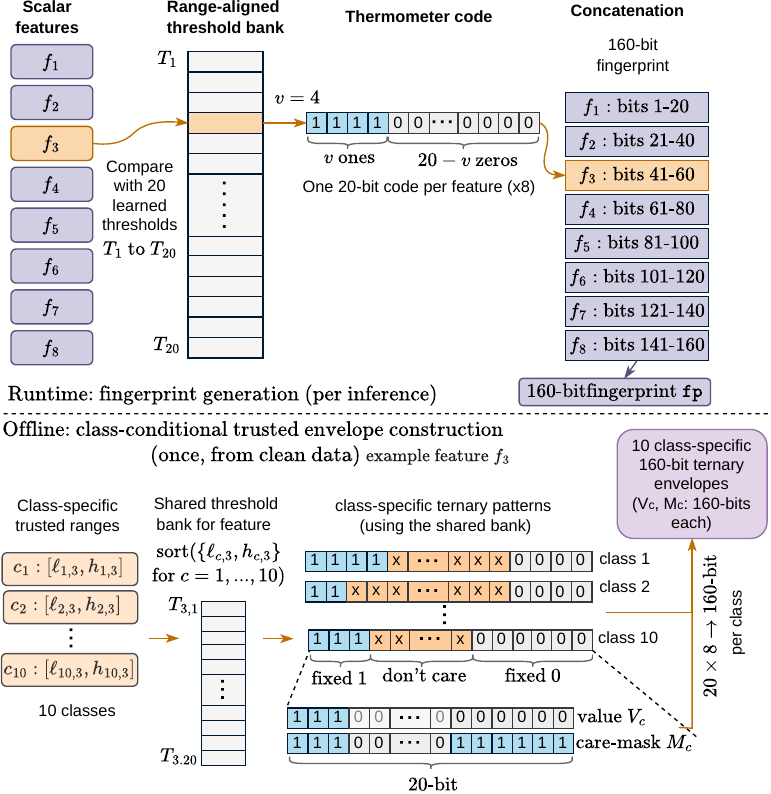}
\caption{
Thermometer encoding and class-conditional envelope construction. Each feature is compared with 20 learned thresholds (top); the eight 20-bit codes form a 160-bit fingerprint. Offline (bottom), class-specific ranges map to ternary value/care-mask envelopes.
\vspace{-2mm}
}
\label{fig:encoding}
\end{figure}

Because the bank contains a threshold corresponding to each lower and upper range boundary, each envelope is exactly representable. For class $c$ andfeature $j$, positions with ${T}_{j,k}<\ell_{c,j}$ are fixed to 1, positions with ${T}_{j,k}\ge h_{c,j}$ are fixed to 0, and those between the bounds are \emph{don't-care}, making a ternary pattern. As registers cannot store a \emph{don't-care}, each entry is stored as a value register $V_c$ and a care-mask register $M_c$, with the mask set at cared positions (Algorithm~\ref{alg:monitor}). Figure~\ref{fig:encoding} shows the construction.

All $C$ entries are compared concurrently. For class $c$, a match occurs iff
\begin{equation}
(\texttt{fp}\oplus V_c)\wedge M_c=\mathbf{0}.
\label{eq:match}
\end{equation}
Of the $C$ parallel match results, TERMon uses the model's predicted class $c$ for the final decision. If there is no match, the inference is flagged. 
%
\begin{algorithm}[t]
\small
\DontPrintSemicolon
\SetKwInOut{Input}{Input}\SetKwInOut{Output}{Output}
\SetKwProg{Fn}{Procedure}{:}{}
\caption{Envelope construction (offline) and runtime decision. $Q_p(\cdot)$
is the empirical $p$-quantile over clean calibration inferences, $\delta$ is
one quantisation step of the fixed-point feature representation, and
$\mathbf{1}[\cdot]$ is the indicator.}
\label{alg:monitor}

\Fn{\textnormal{\textsc{BuildEnvelope}}$(X,\hat{y},C,\alpha)$}{
  \Input{clean features $X\in\mathbb{R}^{N\times d}$, predicted classes
  $\hat{y}\in\{1..C\}^N$, budget $\alpha$}
  \Output{threshold banks $T$, entries $(V_c,M_c)$ for $c=1..C$}

  \ForEach{class $c$}{
    $X^{(c)} \leftarrow$ rows of $X$ with $\hat{y}=c$\;
    \ForEach{feature $j$}{
      $\ell_{c,j} \leftarrow
      Q_{\alpha/2d}\!\left(X^{(c)}_{:,j}\right)$\;
      $h_{c,j} \leftarrow
      Q_{1-\alpha/2d}\!\left(X^{(c)}_{:,j}\right)$\;
    }
  }

  \ForEach{feature $j$}{
    $T_j \leftarrow
    \mathrm{sort}\bigl(
    \ell_{1,j}-\delta,h_{1,j},\ldots,
    \ell_{C,j}-\delta,h_{C,j}\bigr)$\;
    {\scriptsize \tcp*{$2C$ threshold positions; repeated values allowed}}
  }

  \ForEach{class $c$, feature $j$, position $k=1,\dots,2C$}{
    \uIf{$T_{j,k} < \ell_{c,j}$}{
      $V_{c,j,k}\leftarrow 1,\ M_{c,j,k}\leftarrow 1$
      {\scriptsize\tcp*{fixed $1$}}
    }
    \uElseIf{$T_{j,k} < h_{c,j}$}{
      $V_{c,j,k}\leftarrow 0,\ M_{c,j,k}\leftarrow 0$
      {\scriptsize\tcp*{don't-care}}
    }
    \Else{
      $V_{c,j,k}\leftarrow 0,\ M_{c,j,k}\leftarrow 1$
      {\scriptsize\tcp*{fixed $0$}}
    }
  }
  \KwRet{$T,\,V,\,M$}
}

\BlankLine

\Fn{\textnormal{\textsc{Check}}$(x,c,T,V,M)$}{
  \Input{feature vector $x\in\mathbb{R}^d$, predicted class $c$,
  banks $T$, entries $(V,M)$}
  \Output{flag $\in\{0,1\}$}

  \ForEach{feature $j$, position $k$}{
    $\texttt{fp}_{j,k} \leftarrow
    \mathbf{1}[\,x_j>T_{j,k}\,]$
    {\scriptsize\tcp*{comparator output}}
  }

  \ForEach{class $q=1,\dots,C$ \textnormal{ in parallel}}{
    $\mathrm{hit}_q \leftarrow
    \mathbf{1}\!\left[
    (\texttt{fp}\oplus V_q)\wedge M_q=\mathbf{0}
    \right]$\;
  }

  \lIf{$\mathrm{hit}_c=0$}{
    \KwRet{$1$}{\scriptsize\ \texttt{// flagged}}
  }
  \KwRet{$0$}{\scriptsize\ \texttt{// trusted}}
}
\end{algorithm}
%

\subsection{Sequential Decision for Persistent Threats} \label{subsec:sequential_deci}
Persistent threats affect multiple inferences, so TERMon can aggregate flags over time. The prototype uses a two-of-three rule to favor short detection delay, raising an alarm when at least two of the last three inferences are flagged. 
If each inference is independently flagged with probability $p$, the probability that a three-inference window raises an alarm is
\begin{equation}
P_{\mathrm{win}}(p) = 3p^2(1-p) + p^3.
\label{eq:window}
\end{equation}
This aggregation rule is separate from the behavioral envelope and can be adjusted to deployment requirements. Section~\ref{subsec:rq1_fi} examines the resulting trade-off between detection and false alarms.
\section{Reference Detector Methodology} \label{sec:reference_detector}
As the ternary encoding preserves the class-conditional range decision exactly, a missed detection cannot be attributed to the encoding itself. It must instead result from either the decision rule or the monitored features: the range-based rule may be too simple, or the features may not sufficiently distinguish clean and anomalous inferences at the chosen false positive rate. To distinguish them, we evaluate a class-conditional Mahalanobis reference detector~\cite{lee2018mahalanobis} on the same inference data. The reference detector computes a Mahalanobis distance using the monitored features in their original real-valued form and the feature distribution learned for the model's predicted class.
For each predicted class, the reference detector estimates a feature mean and uses a shared shrinkage-regularized covariance matrix for stable covariance estimation. Each inference is compared with the clean feature distribution of its predicted class.
We evaluate the reference detector on both the eight monitored features and larger feature sets. The larger sets add per-channel activation means from the convolutional layers, contributing 32, 64, and 128 additional dimensions, for a maximum of 232. This extended reference tests whether richer activation summaries improve detectability.

The reference detector is diagnostic, not an upper bound on detection performance. Mahalanobis distance is appropriate under a class-conditional Gaussian model, but independent per-feature range checks can perform better when anomalies cause large deviations in only one or a few features. 
The Mahalanobis reference is therefore used to test whether a multivariate decision rule over the same features improves detection. If both detectors perform poorly, the monitored features provide little detection-relevant information, motivating the evaluation of richer feature sets.

\begin{figure*}[ht]
\centering
\includegraphics[width=0.75\linewidth]{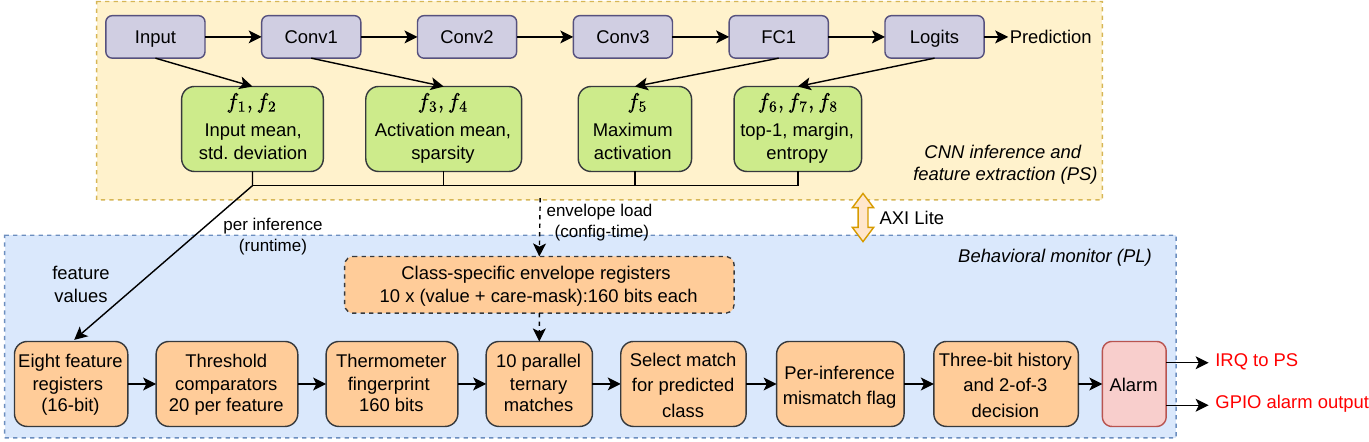}
\caption{The monitor as an AXI4-Lite peripheral alongside the CNN data path on the PYNQ-Z2. Features tapped at four points feed comparator-based thermometer encoding, ten class-specific ternary matches, and the two-of-three sequential rule. Solid arrows show the per-inference runtime datapath and the dashed arrow shows one-time configuration. 
}
\label{fig:arch}
\end{figure*}
%

\section{FPGA Prototype} \label{sec:fpga}
We implement TERMon on a PYNQ-Z2 board with a Zynq-7020 device~(xc7z020clg400-1). The Zynq-7020 system-on-chip combines an ARM Cortex-A9 processing system~(PS) with FPGA programmable logic~(PL), connected through an AXI interconnect. 
\vspace{-2mm}
%
\subsection{Architecture} \label{subsec:arch}
A high-level block diagram of the architecture is shown in Figure~\ref{fig:arch}. In the current prototype, convolutional neural network~(CNN) inference and feature extraction run on the PS, while the trust-decision logic of the monitor, including thermometer encoding, ternary matching, and the sequential rule, is implemented in the PL as a memory-mapped AXI4-Lite peripheral. The PS sends per-inference feature values and predicted class over AXI, and the PL returns the resulting flag and alarm signal. The pipelined implementation requires only two-cycle latency from feature registers to the per-inference flag, excluding the feature extraction on the PS. The same AXI-Lite register interface can also support moving feature extraction into the PL, as it is designed for both the current software-assisted prototype and a fully hardware-integrated implementation.

Eight 16-bit signed feature registers receive the per-inference feature values ($f_1$ to $f_8$) produced at the selected tap points. Each feature is compared against 20 stored thresholds, derived from the class-specific trusted-ranges~(cf.~Section~\ref{subsec:encoding}). The 20 comparator outputs form the thermometer code, and concatenating the eight codes forms a 160-bit fingerprint~($\texttt{fp}$). The fingerprint is compared in parallel with the ten class-specific ternary patterns, each stored as 160-bit \emph{value} register~($V_c$) and \emph{care-mask} register~($M_c$), using $(\texttt{fp}\oplus V)\wedge M$. The result is a 10-bit hit vector, which, together with the predicted class, is used to produce the per-inference flag.

Implementing the ternary envelopes as value and care-mask registers eliminates the need for a dedicated TCAM macro, hence can be implemented using combinational logic and requires no on-chip block RAM~(BRAM). Thresholds and entries together require 5,760 bits and are written once during configuration. The per-inference flag is stored in a flag register and shifted into a three-stage history register that implements the two-of-three rule~(cf. Section~\ref{subsec:sequential_deci}). The resulting alarm is forwarded to the PS as an interrupt request~(IRQ) and to an external safety controller through a general-purpose input/output~(GPIO) pin. 

\subsection{Hardware Implementation Results} \label{subsec:hwresults}
The prototype was synthesized and implemented using Vivado 2025.2 tool and Verilog HDL. Including the AXI-Lite wrapper, TERMon uses only 3,733 LUTs and 6,121 flip-flops, corresponding to 7.02\% and 5.75\% of the device, respectively. It uses no block RAMs and no DSP slices. 
The decision latency is two clock cycles. The achieved clock period is 7.32~ns, corresponding to a maximum operating frequency~($f_{max}$) of 136~MHz. The estimated on-chip power is 151~mW, with dynamic power attributing to only 47~mW. Excluding the AXI-Lite wrapper, the monitor requires only 2,335 LUTs. The implementation cost is dominated by registers, comparators, and ternary match logic rather than memories or dedicated arithmetic units.

\begin{table*}[t]
\small
\centering
\caption{Overview of the evaluated runtime threats. Detection is evaluated at a nominal 1\% false-positive operating point.}
\vspace{-3mm}
\label{tab:threats}
\resizebox{0.93\linewidth}{!}{%
\begin{tabular}{p{2.6cm}p{2.9cm}p{6.2cm}p{4.4cm}}
\toprule
\textbf{Threat category} & \textbf{Dataset / method} & \textbf{Evaluated configurations} & \textbf{Purpose} \\
\midrule
Weight corruption & Random FP32/FP8 weight-bit flips & \{1,10,50,100,200,500\} cumulative flips $\times$ three trajectories (18 corrupted models per format, 36 total) & Persistent model-state corruption \\
\addlinespace
OOD input & CIFAR-10-C & $15$ corruption types $\times$ five severity levels & Distributional change from common corruptions \\
\addlinespace
OOD input & SVHN & Full SVHN test set (cross-dataset) & Cross-dataset distributional shift \\
\addlinespace
Adversarial input & FGSM, PGD-10 & White-box, $\varepsilon=8/255$; detection on the fooled subsets ($3{,}233$ FGSM, $3{,}339$ PGD of $4{,}000$) &
Transient white-box attacks \\
\bottomrule
\end{tabular}
}
\end{table*}
\begin{table*}[t]
\small
\centering
\caption{Overview of the evaluated detectors and representations. All are compared at a $1\%$ false-positive operating point.}
\vspace{-3mm}
\label{tab:detectors}
\resizebox{0.92\linewidth}{!}{%
\begin{tabular}{p{3.0cm}p{3.7cm}p{4.5cm}p{3.6cm}c}
\toprule
\textbf{Detector}   & \textbf{Input representation} & \textbf{Decision rule} & \textbf{Purpose} \\
\midrule
Global range detector & Eight monitored features $f_1$--$f_8$, thermometer-coded &
Flag if any feature leaves its single global trusted range &
Global-range baseline \\
\addlinespace
Class-conditional range detector & Eight monitored features $f_1$--$f_8$, thermometer-coded &
Per-predicted-class trusted ranges (ten entries) &
TERMon hardware design \\
\addlinespace
Mahalanobis reference & Eight monitored features (G8), real-valued &
Class-conditional Mahalanobis distance &
Diagnostic multivariate reference \\
\addlinespace
Extended Mahalanobis & G8 plus per-channel means CH1/CH2/CH3 (up to $232$ dims) &
Class-conditional Mahalanobis distance &
Whether wider activation features help \\
\midrule
Fingerprint-and-Hamming \emph{(abandoned)} & 24-bit SimHash fingerprint &
Hamming distance from five clean centroids &
Design-lessons baseline \\
\bottomrule
\end{tabular}
}
\end{table*}
\section{Evaluation} \label{sec:eval}
\subsection{Experimental Setup} \label{subsec:exp_setup}
We evaluate TERMon primarily on CIFAR-10 using a CNN with FP32 weights and activations. The CNN has three convolutional layers with 32, 64, and 128 channels, followed by two fully connected layers. It contains $\approx$0.6 M parameters and achieves 84.0\% test accuracy. The trusted envelope is class-conditional, giving one ternary entry per CIFAR-10 class, and is fitted on 50,000 clean training inferences with a 1\% false-positive budget split across the eight monitored features, as defined in Section~\ref{sec:monitor_design}. Among 8,000 in-distribution test inferences performed using the uncorrupted model, the measured false-positive rate of the monitor is 1.03\%. For the FP8 weight configuration, the uncorrupted model achieves 83.75\% accuracy with a 0.93\% clean flag rate.

\begin{sloppypar}
Table~\ref{tab:threats} summarizes the evaluated threat instances. We evaluate three threat classes. For fault injection, we generate three cumulative bit-flip trajectories, one per random seed, and evaluate each trajectory after \(n \in \{1,10,50,100,200,500\}\) uniformly random weight-bit flips. We apply the same procedure to both FP32 and FP8~(E4M3) weights, giving 18 corrupted models per format. Because checkpoints within a trajectory share their earlier flips, these 18 models represent three trajectories sampled at six flip counts rather than 18 independent corruption samples. For each corrupted model, we measure classification accuracy and TERMon’s per-inference flag rate on the same 8,000 test inputs. We also tested INT8 and INT4 weights; under random bit flips, both exhibit negligible accuracy degradation even at 500 flips, leaving too few harmful cases to meaningfully assess detection.
For distributional shift, we use the SVHN test set~\cite{netzer2011reading} and CIFAR-10-C corruptions~\cite{hendrycks2019benchmarking}. For adversarial examples, we generate FGSM and PGD-10 attacks at $\varepsilon=8/255$ on 4,000 CIFAR-10 test inputs.
Detection is evaluated on the fooled subsets, containing 3,233 FGSM examples and 3,339 PGD examples, since these are the cases in which the attack successfully changes the model decision. Table~\ref{tab:detectors} summarizes the detectors evaluated and their representations, and the evaluation is presented in the following sections.
\end{sloppypar}
\begin{table}[b]
\centering
\caption{
Fault injection: accuracy drop (percentage points) and per-inference flag rate for corrupted models along three cumulative trajectories.
}
\vspace{-3mm}
\label{tab:fi}
\resizebox{\linewidth}{!}{%
\begin{tabular}{rcccccccccccc}
\toprule
& \multicolumn{4}{c}{\textbf{seed 0}}
& \multicolumn{4}{c}{\textbf{seed 1}}
& \multicolumn{4}{c}{\textbf{seed 2}} \\
\cmidrule(lr){2-5}\cmidrule(lr){6-9}\cmidrule(lr){10-13}
\textbf{flips}
& \multicolumn{2}{c}{FP32} & \multicolumn{2}{c}{FP8}
& \multicolumn{2}{c}{FP32} & \multicolumn{2}{c}{FP8}
& \multicolumn{2}{c}{FP32} & \multicolumn{2}{c}{FP8} \\
\cmidrule(lr){2-3}\cmidrule(lr){4-5}
\cmidrule(lr){6-7}\cmidrule(lr){8-9}
\cmidrule(lr){10-11}\cmidrule(lr){12-13}
& $\Delta$acc & flag & $\Delta$acc & flag
& $\Delta$acc & flag & $\Delta$acc & flag
& $\Delta$acc & flag & $\Delta$acc & flag \\
\midrule
1   & 0.0  & .010 & 0.0  & .009 & 0.0  & .010 & 0.0  & .009 & 0.0  & .010 & 0.0  & .009 \\
10  & 0.0  & .010 & -0.1 & .010 & 72.7 & 1.00 & 0.0  & .009 & 0.0  & .010 & 0.0  & .010 \\
50  & 49.6 & .666 & 0.0  & .010 & 72.7 & 1.00 & 0.0  & .010 & -0.1 & .010 & -0.1 & .010 \\
100 & 49.6 & .665 & 73.8 & 1.00 & 72.7 & 1.00 & 0.0  & .010 & 0.0  & .010 & 0.0  & .010 \\
200 & 49.6 & .666 & -0.1 & .010 & 73.3 & 1.00 & 0.2  & .010 & 0.0  & .011 & 73.8 & 1.00 \\
500 & 61.8 & .814 & 73.8 & 1.00 & 73.3 & 1.00 & 73.8 & 1.00 & 73.7 & 1.00 & 73.8 & 1.00 \\
\bottomrule
\end{tabular}}
\end{table}
%
\subsection{Detection of Weight Faults} \label{subsec:rq1_fi}
Table~\ref{tab:fi} reports the accuracy drop and per-inference detection rate for each corrupted model. The flag rate denotes the percentage of inferences flagged for each corrupted model. For clean or behaviorally benign cases, the flag rate corresponds to the false-positive rate, whereas for harmful corruptions, it represents the per-inference detection rate. Fault count alone does not determine behavioral impact. For the same number of flips, different seeds can produce either negligible or severe accuracy loss. TERMon's flag rate instead tracks the resulting behavioral impact.
The results show a clear separation between benign and harmful corruptions. For FP32, 8 of the 18 corrupted models leave accuracy unchanged, with an accuracy drop of at most 0.1 percentage points and are flagged at 1.0–1.1\%, matching the clean false-positive rate. The remaining ten corrupted models reduce accuracy by 49.6 to 73.7 percentage points and are flagged on 66.5--100\% of inferences, with a mean detection rate of 88.1\%.

This impact-proportional behavior is important for deployment. TERMon reliably detects corruptions that affect model behavior, while maintaining harmless corruptions near the clean false-positive rate. Under the two-of-three rule, the weakest harmful case ($p=0.665$) is detected with at least 99\% probability within twelve inferences, while a healthy model ($p=0.0103$) has a $3.2\times10^{-4}$ false-alarm probability per three-inference window. Because false alarms accumulate over many inferences, a more conservative sequential rule may be preferable. Assuming independent per-inference flags, a four-of-six rule analytically yields an expected false alarm only once per \(9.2\times10^{6}\) healthy inferences while detecting the harmful case with the lowest observed flag rate within twelve inferences with 93.6\% probability. Deployment-specific selection nevertheless requires an explicit time horizon, reset policy, and evaluation on ordered clean traces.
The class-conditional and global envelopes perform almost identically under fault injection. Both achieve a mean detection rate of 88.1\% for harmful corruptions, with clean false-positive rates of 1.03\% and 0.96\%, respectively. This indicates that harmful weight corruptions push monitored features outside the trusted ranges regardless of the predicted class.

We observe the same impact-proportional behavior with FP8 weights. Thirteen of the 18 corrupted models have negligible accuracy change and remain near the clean false-positive rate, whereas the five harmful models lose 73.8 percentage points and are flagged on every inference. All harmful FP8 cases involve bit flips that produce NaN weights.
%
\subsection{Effect of TCAM-Compatible Encoding} \label{subsec:rq2}
\begin{figure}[b]
\centering
\includegraphics[width=0.85\columnwidth]{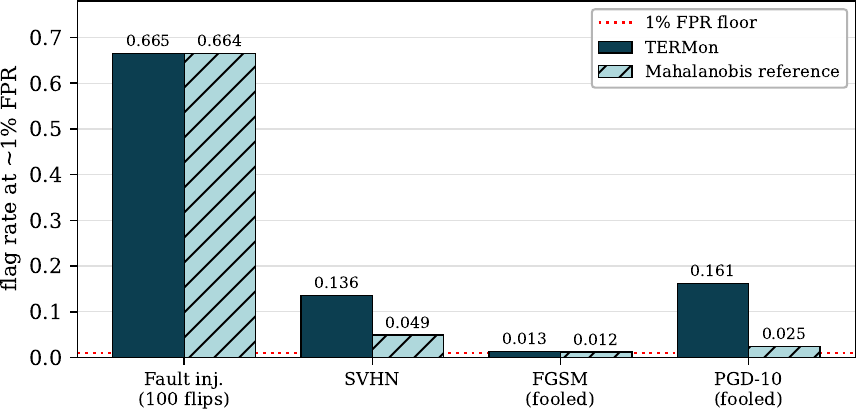}
\caption{
Per-inference detection rate: TCAM range detector vs. continuous Mahalanobis reference detector.
}
\label{fig:tcam_vs_ref_det}
\end{figure}

We first evaluate whether the ternary encoding introduces any loss relative to the unquantized range test. Across the clean test set, the OOD set, and all evaluated corrupted models, the ternary and unquantized range tests align on every inference. Thus, any missed detection is due to the decision rule or the monitored features rather than the ternary encoding.
Figure~\ref{fig:tcam_vs_ref_det} compares TERMon with the Mahalanobis reference at matched false-positive operating points. TERMon has a measured false-positive rate of 1.03\%, and the reference detector is thresholded at 1\%. For heavy fault injection, TERMon achieves 66.5\% per-inference detection, compared with 66.4\% for the reference detector. Independent per-feature ranges therefore perform similarly to a multivariate Mahalanobis detector over the same features.

On the remaining threats TERMon exceeds the reference. It detects 13.6\% of SVHN inputs against 4.9\%, and 16.1\% of successful PGD examples against 2.5\%. Per-feature range checks can therefore outperform the reference detector when an anomaly causes one or more individual features to leave their normal ranges, as discussed in Section~\ref{sec:reference_detector}. Both rates nevertheless remain too low for reliable per-input OOD detection. On successful FGSM examples, both detectors remain close to the false-positive floor, indicating that the monitored features carry little adversarial-example information.

%
\subsection{Limits Under OOD and Adversarial Inputs} \label{subsec:limits}
The results show that reliable per-input OOD detection is not achieved for this model and feature set at a 1\% false-positive operating point. The reference detector identifies only 4.9\% of SVHN inputs using the eight monitored features. Adding per-channel activation means from all three convolutional layers increases the feature dimension to 232, but improves detection only to 5.4\%. The feature distributions explain this limited gain, as some clean CIFAR-10 inputs are hard examples that cause the model to produce low confidence predictions with high entropy. Their behavioral signatures overlap with those of SVHN inputs. At the 99th percentile of the clean distribution, the monitored statistics therefore do not separate these populations well.

\begin{figure}[t]
\centering
\includegraphics[width=0.8\columnwidth]{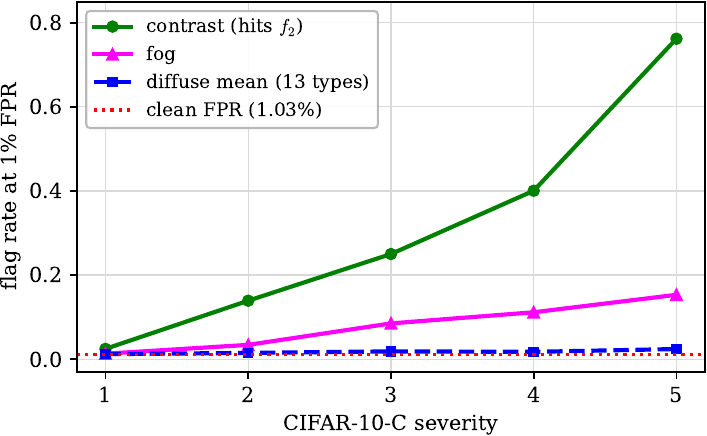}
\caption{
CIFAR-10-C flag rate at a 1\% false-positive operating point (flag rate here is the detection rate, as all inputs are corrupted).
}
\label{fig:c10c}
\end{figure}

CIFAR-10-C corruptions show the same pattern, with contrast corruption as the main exception, as shown in Figure~\ref{fig:c10c}. At severity 5, the detector flags 76.2\% of contrast-corrupted inputs, while most other corruptions remain far lower. For example, fog reaches 15.3\%, brightness 5.7\%, impulse noise 5.1\%, snow 4.1\%, and every other corruption stays below 2.3\%. Contrast is detectable as its effect is concentrated in one monitored statistic, the input standard deviation $f_2$. Section~\ref{subsec:lessons} shows that thresholding $f_2$ alone detects 82.4\% of contrast-corrupted inputs, whereas more diffuse corruptions such as noise, blur, and compression spread their effect across features and are not reliably detected at the same false-positive rate.

Because distributional drift can persist over many inferences, sequential aggregation can still be useful even when per-input detection is weak. TERMon reaches 13.6\% per-inference detection on SVHN against a 1.03\% false-positive rate, which supports windowed detection of sustained shift. We nevertheless make no claim of reliable per-input OOD detection. 
Adversarial inputs behave differently under different envelopes. On successful PGD examples TERMon flags 16.1\% of inferences, against 4.1\% for a single global envelope and 2.5\% for the reference detector, due to the class-conditional envelope. A successful attack changes the predicted class, so TERMon checks the inference against the trusted range of that class. This can expose behavior that still appears normal under a global range. The effect does not appear on FGSM, where TERMon flags only 1.3\%. 
Since most attacks still pass, we do not claim adversarial robustness. 
The differing FGSM and PGD rates indicate attack-dependent detectability, but neither provides reliable per-input adversarial detection at the 1\% false-positive operating point. The poor Mahalanobis results further show that a multivariate decision rule over the same features does not resolve this limitation, motivating complementary features or detectors. This limited per-input adversarial detectability is consistent with prior work showing that adaptive attacks can bypass adversarial-example detectors~\cite{carlini2017detected}. 
Reliable per-input OOD detection should likewise be treated as complementary to TERMon’s temporal monitoring of persistent shifts. 
%
\subsection{Lessons from Alternative Designs} \label{subsec:lessons}
The final design of TERMon is motivated by measured failures of alternative representations, aggregation rules, and feature sets. 

\textit{1. Random-projection encodings discard the relevant information.} 
Our initial design used random hyperplane projections, similar to SimHash~\cite{charikar2002simhash}, to map feature vectors to short binary fingerprints so that similar vectors map to similar bit patterns. Because this encoding preserves direction but discards magnitude, it loses information when faults change activation magnitudes without strongly changing their overall pattern. Consequently, SVHN inputs were flagged at only 0.27\%, below the 0.71\% clean false-positive rate of the SimHash detector.

\textit{2. Short Hamming envelopes can become too permissive.} A modified variant of SimHash detector kept the fingerprint but replaced random projection with a learned 24-bit encoding. It defined the trusted region as a Hamming ball around five clean centroids. An inference is trusted if its fingerprint lies within a fixed Hamming radius of some centroid. With 24-bit fingerprints, the Hamming radius needed to keep clean false positives low was 10 bits. A single Hamming ball of radius 10 covers $\sum_{k=0}^{10}\binom{24}{k}/2^{24}=27.1\%$ of all possible fingerprints. With five stored centroids, their combined accepted region expands further, leaving little room to distinguish anomalous behavior, while reducing the radius rejects more clean fingerprints. Hence, no radius provided useful separation in this encoding.

\textit{3. Summed-distance aggregation dilutes localized anomalies.} Some anomalies are concentrated in one or a few features. For severe contrast corruption, the input standard deviation alone detects 82.4\% of inputs at a 1\% false-positive rate, compared with 27.6\% for the eight-feature Mahalanobis reference and 0.1\% for summed-Hamming matching. TERMon therefore checks each feature independently and flags an inference when any feature leaves its trusted range.

\textit{4. Aggregate ROC metrics can hide poor performance at strict false-positive rates.} The output-entropy feature achieves an area under the receiver operating characteristic curve (AUROC) of 0.836 when distinguishing clean and SVHN inputs, but detects only 2.8\% of SVHN at a 1\% false-positive rate. Candidate features must therefore be evaluated at the intended operating point rather than by aggregate metrics alone.

\textit{5. More feature dimensions do not necessarily improve detection.}
Under heavy fault injection, the Mahalanobis reference detects 66.4\% using the eight monitored features but only 1.3\% using the 128 per-channel means from the final convolutional layer. This indicates that many weakly affected dimensions can dilute the effect of a few informative features, motivating a small set of features chosen by both statistic and tap point.

\textit{6. Thresholds must align with the decision boundaries.}
An earlier encoding compared each monitored feature against 15 quantile thresholds $Q_{k/16}$, $k=1,\ldots,15$, corresponding to 16 equal-probability bins. These thresholds span the 6.25th to 93.75th percentiles, whereas with $\alpha=1\%$ and $d=8$ the trusted bounds lie at the 0.0625th and 99.9375th percentiles. All 15 thresholds therefore fall inside the trusted range, making every ternary position don't-care and the pattern accept every fingerprint.
Avoiding this would require approximately 1600 bins per feature just to place a threshold at the lower bound. Equally spaced thresholds performed poorly for the opposite reason, as rounding the bounds outward to the nearest grid points widened the trusted range. For a corruption that reduces model accuracy by 49.6 percentage points, detection falls from 66.6\% with range-aligned thresholds to 2.0\% with equally spaced thresholds. TERMon therefore uses the trusted-range boundaries themselves as thresholds.

Together, these results motivate TERMon's selected scalar features, per-feature range checks, and range-aligned threshold construction. 
More generally, candidate features and representations should be evaluated against the target threat at the required false-positive rate before being committed to hardware.

\section{Related Work and Comparison} \label{sec:related_work}
\begin{sloppypar}
Our work builds on prior observations that weight faults often appear as abnormal activation values, but differs from existing defenses in how the abnormal behavior is detected and how broadly the mechanism is evaluated. Table~\ref{tab:positioning} separates what TERMon evaluates from what the results support, making explicit that the monitor is effective for behavior-changing faults, limited for OOD detection, and not claimed as an adversarial defense. We then compare against activation-range defenses, weight-integrity checks, interval-based monitors, and hardware-level fault-tolerance mechanisms. Table~\ref{tab:defense_comparison} compares the same body of work by detection mechanism.
\end{sloppypar}

\textit{Activation-range defenses for weight faults.}
The closest prior work detects or mitigates weight faults through their effect on activation magnitudes. Hong et al.~\cite{hong2019terminal} show that bit flips in FP32 weights can cause severe accuracy degradation, specifically with exponent-bit flips, while many mantissa-bit flips have little effect. Building on this observation, Ranger~\cite{chen2020ranger} and its non-uniform extension~\cite{geissler2021range} insert range-restriction layers that clip out-of-range activations. TERMon relies on a similar observation that harmful weight corruptions tend to produce unusually large internal activation values, which also explains the impact-proportional behavior observed in our evaluation. The mechanism, however, is different. Ranger-style defenses modify the model dataflow and mitigate faults by clipping activations, whereas TERMon observes selected scalar behavioral features, raises a hardware flag, and does not alter the model computation.
%
\begin{table}[b]
\small
\setlength{\tabcolsep}{4pt}
\centering
\caption{Positioning relative to prior runtime defenses by threat coverage. H~=~hardware implementation; DL~=~deterministic latency.}
\vspace{-3mm}
\label{tab:positioning}
\resizebox{\linewidth}{!}{%
\begin{tabular}{lccccc}
\toprule
\textbf{Approach} & \textbf{Faults} & \textbf{OOD} & \textbf{Adv.} & \textbf{H} & \textbf{DL} \\
\midrule
Mahalanobis OOD detector~\cite{lee2018mahalanobis}             & --         & \checkmark & \checkmark & --         & --         \\
Feature squeezing~\cite{xu2017feature}                         & --         & --         & \checkmark & --         & --         \\
Fault-tolerant accelerators~\cite{zhang2018faultaware,fsa2023} & \checkmark & --         & --         & \checkmark & \checkmark \\
Weight-integrity FI detection~\cite{javaheripi2021hashtag,li2021radar} & \checkmark & --   & --         & \checkmark & \checkmark \\
\midrule
\textbf{TERMon} (evaluated) & \checkmark & \checkmark & \checkmark & \checkmark & \checkmark \\
\textbf{TERMon} (effective) & \checkmark & limited    & no claim   & \checkmark  & \checkmark \\
\bottomrule
\end{tabular}
}
\vspace{-2mm}
\parbox{\linewidth}{\footnotesize{\emph{evaluated} row marks the threat classes for which we report experiments and \emph{effective} row reports the outcome~(\emph{limited}: weak per-input detection; \emph{no claim}: no robustness claimed).}}
\end{table}
%
\begin{table*}[t]
\small
\centering
\caption{Comparison of runtime defense paradigms for ML accelerators by
detection mechanism.}
\vspace{-3mm}
\label{tab:defense_comparison}
\resizebox{0.95\linewidth}{!}{%
\begin{tabular}{p{3.4cm}p{2.7cm}p{4.0cm}p{3.6cm}p{3.0cm}}
\toprule
\textbf{Defense} & \textbf{Approach type} & \textbf{Detection basis} & \textbf{Execution overhead} & \textbf{Invasiveness} \\
\midrule
Mahalanobis OOD~\cite{lee2018mahalanobis} & Software reference & Class-conditional distance in continuous feature space & High (FP distance, covariance storage) & None (post-inference software) \\
\addlinespace
Outside the Box~\cite{henzinger2020outside} & Software abstraction & Interval (box) membership over neuron valuations & Low in software (FP membership test) & None (reads hidden layers) \\
\addlinespace
Range supervision / Ranger~\cite{chen2020ranger, geissler2021range} & In-graph restriction & Out-of-bound activation magnitudes (full tensors) & Per-layer, in dataflow & Medium (inserts protection layers) \\
\addlinespace
UniGuard~\cite{uniguard2024} & Physical / side-channel & Power-trace anomalies (TDC voltage sensor) & Small sensor; off-chip ML inference & Low (external sensor, black-box model) \\
\addlinespace
Fault-tolerant accelerators~\cite{zhang2018faultaware,fsa2023} & PE redundancy & Faulty-PE detection (BIST) for permanent faults & Re-compute / bypass logic per PE & Medium (alters systolic array) \\
\addlinespace
\textbf{TERMon (this work)} & \textbf{Hardware peripheral} & \textbf{Class-conditional scalar ranges} & \textbf{Deterministic two-cycle} & \textbf{Low (non-intrusive tap)} \\
\bottomrule
\end{tabular}
}
\end{table*}

\textit{Weight-integrity defenses.}
Several defenses detect weight fault injection by verifying weight integrity directly. RADAR~\cite{li2021radar} derives checksum-based signatures over interleaved and masked weight groups, compares them against golden signatures at run time, and zeroes the flagged groups to recover accuracy. HASHTAG~\cite{javaheripi2021hashtag,javaheripi2022acchashtag} hashes weights in vulnerable layers and validates the hashes during inference, with provable detection bounds. These methods are lightweight and effective against targeted bit-flip attacks, but they verify weight integrity rather than inference behavior. They therefore flag detected weight changes regardless of behavioral impact and do not address threats that leave weights intact, such as sensor drift, OOD inputs, or adversarial inputs. TERMon instead checks observed inference behavior. For weight corruption, this makes detection impact-proportional, as it responds to behavioral impact rather than to weight modification alone. The same behavioral feature interface is also used to evaluate, and expose the limits of, detection under distributional shift and adversarial inputs.

\textit{Interval-based behavioral monitoring.}
Outside the Box~\cite{henzinger2020outside} builds interval regions from clean neuron activations and flags inputs whose activations leave those intervals. TERMon uses a similar range-based idea, but targets a different implementation point. This is conceptually close to our use of trusted ranges, but the implementation target is different. Instead of software checks over neuron-level activations, TERMon reduces behavior to eight scalar features and maps their trusted ranges to thermometer-encoded ternary patterns. This turns the range check into a lightweight hardware match using value and care-mask registers, without a dedicated TCAM macro.

\textit{Hardware-level anomaly and fault-tolerance mechanisms.}
Some hardware defenses monitor or harden the execution substrate rather than the neural-network behavior. UniGuard~\cite{uniguard2024}, for example, detects power side-channel anomalies using a time-to-digital converter voltage sensor and an off-chip classifier. Other works improve hardware resilience by adding redundancy at the processing-element level~\cite{fsa2023} or by using fault-aware pruning to tolerate permanent defects~\cite{zhang2018faultaware}. These approaches are complementary to TERMon, as they target physical side channels, hardware faults, or permanent defects rather than deviations in inference behavior.
\vspace{-2mm}
\section{Discussion and Future Work} \label{sec:discussion}
TERMon is complementary to weight-integrity checks rather than a replacement for them. Integrity checks suit settings where recovery is cheap, whereas a behavioral monitor suits settings where the response is expensive. This is particularly relevant to applications such as remote sensors and satellites. In this section, we discuss the limitations and future research directions.

\textit{Limitations.}
TERMon does not provide adversarial robustness and has limited detection when OOD or adversarial inputs keep the monitored features within their trusted ranges. Our threat model assumes that the monitor cannot be modified or disabled and excludes attacks specifically crafted to keep the monitored features within range. 
Finally, the evaluation uses a single CIFAR-10 CNN and random FP32/FP8 (E4M3) weight faults; broader validation across architectures, targeted bit-flip attacks, and other weight formats remains future work.

\textit{Future work.}
TERMon’s small footprint enables multi-model extensions. A shared monitoring datapath could be time-multiplexed across models with model-specific thresholds and envelopes, while a TCAM macro could replace registers when many envelopes must be stored. Threat coverage should be strengthened through protected or keyed thresholds, targeted attacks on low- and mixed-precision models, and adaptive attacks that jointly optimize against the classifier and monitored features. Deployment-aware sequential rules should also consider explicit time horizons and reset policies. Finally, moving feature extraction from the Zynq PS into dedicated logic would complete the hardware path and enable end-to-end evaluation of trust, area, latency, power, and resource overhead.
%
\section{Conclusion} \label{sec:conclusion}
This paper presented TERMon, a hardware runtime monitor for detecting behavioral anomalies in edge-AI inference. TERMon checks eight features against ranges learned from clean inferences and maps the check to a thermometer-encoded ternary envelope. The evaluation shows that this mechanism is effective for persistent, behavior-changing weight faults. Harmful corruptions are detected on 66.5--100\% of inferences, with an average detection rate of 88.1\%, while benign corruptions remain near the clean-inference false-positive rate. The ternary encoding produces identical detection results to the corresponding unquantized range test, showing that the ternary representation does not reduce the range detector's performance for the evaluated fault setting. We also observe that out-of-distribution and successful adversarial inputs are largely invisible in the monitored features at a strict false-positive operating point. The FPGA prototype implements the monitoring core using 3,733 LUTs, no BRAMs, no DSPs, with two-cycle latency. TERMon therefore provides a low-cost behavioral guardrail for runtime failures that are visible in selected inference features.
\bibliographystyle{ACM-Reference-Format}
\bibliography{TERMon_camera_ready}

\section*{Use of Generative AI}
Generative AI tools were used in the preparation of this work. AI tools were used as a writing assistant to improve the grammar, formatting, and clarity of the writing. They were also used to generate and optimize parts of the source code, such as the reference detector, the feature extractor, the fault-injection experiments, and the scripts used to produce the evaluation plots. The overall research direction, the system design, the hardware implementation, and the interpretation of results were carried out by the authors. The authors reviewed, tested, and verified all AI-assisted text and code, and take full responsibility for the entire content of this work.
\end{document}